\documentclass[11pt, oneside, onecolumn]{article}   	% use "amsart" instead of "article" for AMSLaTeX format
\usepackage{geometry}                		% See geometry.pdf to learn the layout options. There are lots.
\usepackage{graphicx}				% Use pdf, png, jpg, or eps with pdflatex; use eps in DVI mode
\usepackage{amssymb,amsmath}
\usepackage{gensymb} 
\usepackage{hyperref} 
\hypersetup{	
colorlinks=true, 
linkcolor= blue,	
citecolor=blue,	
} 
\usepackage{subfigure, color}
\usepackage[dvipsnames]{xcolor} % to be able to use ~68 colors instead of ~19 default latex colors
\usepackage{tikz}

\usepackage{cite}
\usepackage{gensymb} %degree symbol
\usepackage{subfigure, color, xcolor,bm}

\def\f{\mathbf{f}}
\def\r{\mathbf{r}}

\def\i{{\rm i}}
\def\a{{\rm a}}
\def\b{{\rm b}}

\def\c{{\rm c}}
\def\xbi{x_{\rm b}^{({\rm i})}}
\def\xbj{x_{\rm b}^{({\rm j})}}

\title{Basal coupling modulates bifilament synchronization}
\title{Intracellular coupling modulates biflagellar synchrony}
\author{Hanliang Guo$^{1,2}$, Yi Man$^1$, Kirsty Y. Wan$^{3\,*}$, Eva Kanso$^{1}$\footnote{Corresponding authors: K.y.wan2@exeter.ac.uk and Kanso@usc.edu} \\[1ex]
\small{${}^1$ Aerospace \& Mechanical Engineering, University of Southern California, Los Angeles, CA 90089, USA} \\
\small{${}^2$  Mathematics, University of Michigan, Ann Arbor, MI 48109, USA} \\
\small{${}^3$ Living Systems Institute, University of Exeter, Exeter, EX4 4QD, UK}}

\date{}							

\begin{document}

\maketitle

\begin{abstract}
\small{Beating flagella exhibit a variety of synchronization modes.  This synchrony has long been attributed to hydrodynamic coupling between the flagella. However, recent work with flagellated algae indicates that a mechanism internal to the cell, through the contractile fibres connecting the flagella basal bodies, must be at play to actively modulate flagellar synchrony. Exactly how basal coupling mediates flagellar coordination remains unclear. Here, we examine the role of basal coupling in the synchronization of the model biflagellate \textit{Chlamydomonas reinhardtii} using a series of mathematical models of decreasing level of complexity.  We report that basal coupling is sufficient to achieve inphase, antiphase, and bistable synchrony, even in the absence of hydrodynamic coupling and flagellar compliance. These modes can be reached by modulating  the activity level of the individual flagella or the strength of the basal coupling. We observe a slip mode when allowing for differential flagellar activity, just as in experiments with live cells. We introduce a dimensionless  ratio of flagellar activity to basal coupling that is predictive of the mode of synchrony. This ratio allows us to query biological parameters which are not yet directly measurable experimentally. Our work shows a concrete route for cells to actively control the synchronization of their flagella.}
\end{abstract}

\section*{Introduction}
\label{sec:intro}

Cilia and flagella often exhibit synchronized behaviour; this includes phase locking, as seen in the alga \textit{Chlamydomonas  reinhardtii}~\cite{Ruffer1985,Ruffer1987}, and metachronal wave formation in the respiratory cilia of higher organisms~\cite{sanderson1981}. 
Since the observations by Gray and Rothschild of phase synchrony in nearby swimming spermatozoa~\cite{Gray1928, Rothschild1949}, it has been a working hypothesis that synchrony arises from hydrodynamic interactions between beating filaments~\cite{Guirao2007, Niedermayer2008, Golestanian2011, Uchida2011, Uchida2012, Brumley2012}. 
Recent work on the interaction dynamics of physically separated pairs of flagella isolated from the multicellular alga Volvox has shown that hydrodynamic coupling alone is sufficient to produce synchrony in some cases~\cite{Brumley2014,Wan2014}. 
These observations were reproduced experimentally with oscillating bead models~\cite{Kotar2010, Bruot2012} and \textit{in-silico} in the context of hydrodynamically-coupled filaments~\cite{Guo2018, Man2020b}.
However, in many unicellular organisms flagellar synchrony seems to be more complex:
recent work with flagellated algal cells indicates that a mechanism internal to the cell must be at play in the active control of flagellar synchrony~\cite{Quaranta2015, Wan2016}.

%---------------------
\begin{figure*}[!t]
        \centerline{\includegraphics[scale=.9]{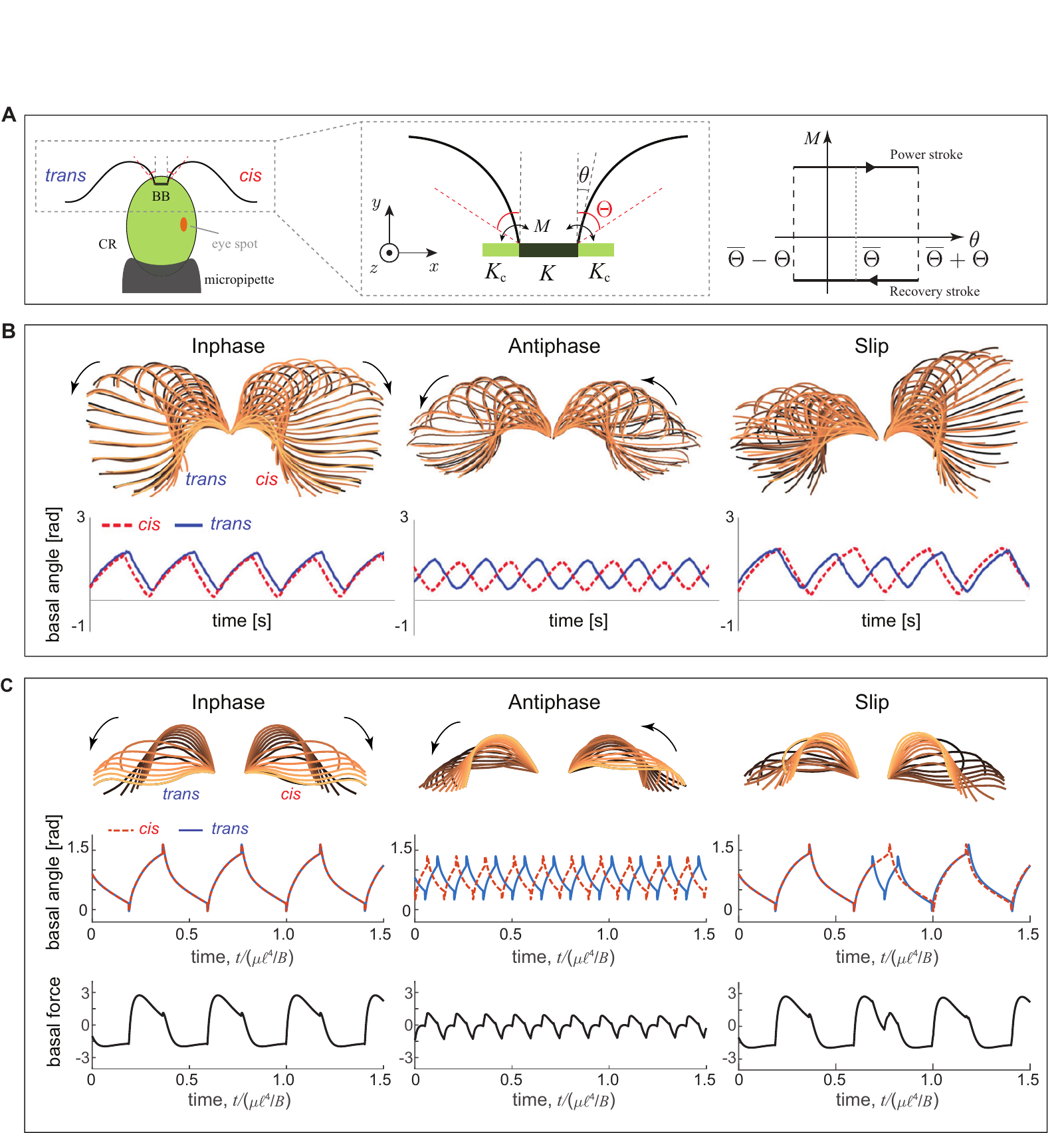}}
	\caption[]{\footnotesize (A) Schematic of the biflagellate alga \textit{Chlamydomonas reinhardtii} (CR), held in place using a micropipette as in~\cite{Wan2016}, with \textit{cis} and \textit{trans} flagella connected via basal fibres (BF); the flagella are modeled as a pair of filaments  coupled at their base via elastic springs. The filaments are driven into oscillations by an active moment $M$ at the filament base that switches direction when the basal angle $\theta$ reaches pre-defined locations $\pm\Theta$ relative to the average basal angle $\bar{\Theta}$, akin to the geometric switch model~\cite{Kotar2010,Bruot2012, Guo2018}. (B) Experimental data: snapshots of the flagellar waveforms during one oscillation cycle and time evolution of basal angles for {inphase}, {antiphase} and slip motions. Data reproduced from~\cite[Figure 3]{Wan2014}. Five consecutive beats are shown with time being color-coded.  (C) Filament pair model: snapshots of filament waveforms during one beating period and time evolution of basal angles  as well as the respective basal spring force showing inphase ($\Theta= 0.2\pi$), antiphase ($\Theta= 0.1\pi$), and phase-slip. The bases of the flagella are fixed in the snapshots for aesthetic purposes.  Model parameters and filament simulations are provided in the electronic supplemental material. 
	} 
	\label{fig:Chlamy}
\end{figure*}
%------------------------

\textit{Chlamydomonas reinhardtii} has emerged as a model system for probing dynamic flagellar synchrony~\cite{Pedley1992, Goldstein2015}. A single eyespot breaks the cell's bilateral symmetry, distinguishing the \textit{cis} flagellum (closer to the eyespot) from the \textit{trans} flagellum (Figure~\ref{fig:Chlamy}A).  The \textit{cis} and  \textit{trans} flagella are active filaments actuated by internal molecular motor proteins acting on an intricate structure of microtubules known as the ciliary axoneme~\cite{Satir1990, Satir2007}. Each beat cycle comprises a power stroke, which generates forward propulsion, and a recovery stroke, in which the flagella undergo greater curvature excursions, thereby overcoming the reversibility of Stokes flows~\cite{Purcell1977, Lauga2009}. 
The basal bodies (BB) from which the flagella nucleate are not essential for flagellar beating; isolated axonemes (detached from the cell body) continue to beat when reactivated in ATP~\cite{Bessen1980,  Mukundan2014}. However, the contractility of inter-BB connections in the algal flagellar apparatus \cite{Salisbury1989} contributes to flagellar coordination~\cite{Wan2016}. 
Wild type cells with intact basal connections swim with a familiar {inphase} breaststroke, with flagella that beat inphase in opposite directions interrupted occasionally by extra beats (`slips') of the \textit{trans} flagellum. 
During phototaxis, the \textit{cis} and \textit{trans} flagella are thought to respond differentially to elevations in the intracellular calcium levels \cite{Kamiya1984}, thus altering both the flagellar beat waveform and synchrony.  Transient loss of biflagellar synchrony occurs stochastically at rates sensitive to light and other environmental factors~\cite{Ruffer1987,Ruffer1998,Josef2005,Wan2014}. This stochastic switching between synchronous {inphase} with slips and asynchronous beating is similar to the run-and-tumble motion of bacteria, with sharp turns taking the place of tumbles~\cite{Polin2009}.

In a mutant (\textit{vfl3)} with impaired basal connections, % and variable number of flagella, 
synchrony is almost completely disrupted~\cite{Wan2016}. 
A different mutant (\textit{ptx1}), which is deficient in phototaxis and regulation of flagellar dominance, exhibits stochastic switching between {inphase} and {antiphase} modes in a way reminiscent of the synchronous or asynchronous transitions of the wild type~\cite{Leptos2013,Wan2014}.
The {antiphase} is characterized by flagella that beat in the same direction, with attenuated beat amplitude and increased beat frequency~\cite{Leptos2013}. 
Further, the flagellar waveform in the {antiphase} mode carries striking similarity to that of the flagellum which accumulates additional cycles during a phase slip of the wild type~\cite{Wan2014}. 
In Figure~\ref{fig:Chlamy}B, we reproduce experimental results from~\cite[Figure 3]{Wan2014} with overlaid sequences of tracked flagella showing a normal {inphase} of a wild type cell, the {antiphase} gait of the phototaxis mutant, and a stochastic slip event in which the  wild type \textit{trans} flagellum transiently executes an extra beat with an attenuated beat amplitude. 

Taken together, multiple lines of evidence from \textit{Chlamydomonas} (namely, the above observations of wild type flagellar synchrony, the mutant with impaired basal connections, and the phototaxis mutant) as well as from other algal species \cite{Wan2020} strongly imply that intracellular coupling mediates flagellar synchrony. 
Although the precise biochemical and biophysical mechanisms by which intracellular activity regulates flagellar coordination are yet to be determined, a major working hypothesis is that this is achieved through contractility of the flagellar basal apparatus.
The observation of spontaneous transitions between extended inphase and antiphase beating in the phototaxis mutant suggests that multiple synchronization states might be achievable through changes in the mechanical properties of the basal fibres, which couple the flagella basal bodies at specific locations. 
The contractility of these centrin-based fibres is well-established, as is their sensitivity to intracellular calcium concentration \cite{Salisbury1989}. 
Thus the flagella pair forms a biophysical equivalent to Huygens' clocks. 
Two oscillating pendula tend toward synchrony (or antisynchrony) when attached to a common support, whose flexibility provides the necessary mechanical coupling.

{With mounting evidence for intracellular coupling in the \textit{Chlamydomonas} system, 
several models have emerged recently for how this might work \cite{Klindt2017,Liu2018}.
In~\cite{Klindt2017}, experimentally-derived flagellar beat patterns were mapped to a limit-cycle oscillator, to construct a minimal model for basal spring coupling. Both flagellar waveform compliance and basal coupling were found to stabilize antiphase synchrony when acting in isolation, while their superposition could stabilize inphase synchrony.
Meanwhile, in \cite{Liu2018}, the authors extended a classical bead model by coupling the two flagella with a non-isotropic elastic spring, and showed that the synchronization mode depended on the relative stiffness of the spring in two orthogonal directions; altering spring stiffness could lead to transitions between inphase and antiphase. 
The inclusion of hydrodynamic interactions made the two beads more likely to synchronize antiphase.
These studies present elegant physical models and provide significant insights into the interplay between basal coupling and hydrodynamic interactions for flagella synchronization. However, neither model explicitly accounts for the marked change in the beating patterns reported \textit{in vivo}~\cite{Wan2014}.}

Here, we propose a new theoretical model for flagellar synchronization via basal coupling, motivated by our ongoing work on flagellar dynamics~\cite{Ling2018,Guo2018,Man2019,Man2020a,Man2020b}. Specifically, we develop an elasto-hydrodynamic filament model in conjunction with numerical simulations to demonstrate that it is possible for a pair of filaments to reach multiple synchronization states simply by varying the intrinsic filament activity and the strength of elastic basal coupling between the two filaments. 
We then derive a minimal model where each flagellum is represented by a rigid dumbbell, to better understand the physical processes driving the synchronization dynamics. 
{The goal of the filament model is to capture many of the details of the observed flagellar waveforms and to show that, with these detailed ciliary waveforms, varying either the degree of filament activity or the strength of basal coupling, is sufficient to transition between inphase and antiphase synchrony. The dumbbell model, because of its simplicity and minimal ingredients, serves to identify the essential ingredient that causes different modes of synchrony.}
We find that a single dimensionless number, defined to be the relative strength of internal flagellar actuation to strength of the basal coupling, could readily predict the pairwise synchronization mode. 
These findings provide new insights into an emerging class of intracellular flagellar coupling mechanisms -- distinct from hydrodynamic interactions -- which may be responsible for flagellar coordination in some unicellular organisms. 
In contrast to {the previous} models of basal coupling \cite{Klindt2017,Liu2018}, our approach accounts for significant features of flagellar beat patterns, which have been shown experimentally to emerge in association with distinct synchronization state.
Finally, we discuss the biological implications of our findings for gait selection and control in flagellates.

\section*{Filament model}

Figure~\ref{fig:Chlamy}(A) summarises our model of the two flagella of \textit{Chlamydomonas} coupled
through basal body interactions. Following previous theoretical work~\cite{Guo2018} and experimental
measurements of flagellar waveform~\cite{Wan2014},  we represent the two flagella by two filaments, each of length $\ell$ and radius $a$, immersed in a fluid of viscosity $\mu$, with each filament driven at its base by a moment $M$. The moment $M$ is a configuration-dependent moment that switches direction when the basal angle of the filament, denoted $\theta$, reaches predefined values $\pm \Theta$ relative to the average angle $\bar{\Theta}$. This actuation mechanism is reminiscent of the geometric switch model for colloidal systems studied in~\cite{Kotar2010, Bruot2012}, yet simpler than the geometric switch actuation along the entire filament length used in~\cite{Buchmann2017}. Here, the filament activity is represented by two parameters: the active moment $M$ and amplitude $\Theta$.
Intracellular connections, through striated fibres joining the basal bodies of the two flagella, are modeled by an elastic spring of stiffness $K$ that couples the two filaments directly at their bases. 
We also allow for couplings between each basal body and the surrounding cytoskeletal structures, these are accounted for by two additional elastic springs each of stiffness $K_\c$ which connect the filaments to a fixed location along the $x$-direction (Figure~\ref{fig:Chlamy}A). The spring $K_c$ loosely anchors the filament and prevents it from drifting sideways.

We let $\r(s,t)$ denote the  position of one of the filaments as a function of time $t$ and arc length $s$, and the subscripts $(\cdot)_t$ and $(\cdot)_s$ denote differentiation with respect to $t$ and $s$, respectively. We express the components of $\r(s,t)$ in a fixed inertial frame $\{\mathbf{e}_x,\mathbf{e}_y,\mathbf{e}_z\}$, with $\mathbf{e}_x$ the unit vector along the direction of the basal connection and $\mathbf{e}_z$ the direction orthogonal to the plane of filament deformation (Figure~\ref{fig:Chlamy}A). We let $x_\b(t)$ denote the position of the filament base, measured from the equilibrium configuration of the basal springs. Motivated by~\cite{Sartori2016}, we allow the filament to have a  configuration-dependent reference curvature given by the curvature vector $\boldsymbol{\kappa}_{\rm o} = (\r_{ss}\times\r_{s})_{\rm{o}}$, where the subscript ${(\cdot)}_{\rm o}$ refers to a non-straight reference configuration. For naturally straight filaments, the reference curvature is identically zero. The internal elastic moment $\mathbf{M}$ is related to the filament current and reference curvature via the Hookean constitutive relation $\mathbf{M} = B(\boldsymbol{\kappa} - \boldsymbol{\kappa}_{\rm o})$, where $B$ is the filament bending rigidity. 
The internal force $\mathbf{N}(s,t)$ along the filament can be decomposed into a tangential component (tension) which enforces filament inextensibility and a normal component that includes the filament resistance to bending. We let $\f(s,t)$ denote the hydrodynamic force density (density per unit length) exerted by the filament on the fluid.
Balance of forces and moments on each filament $(\i=1,2)$, together with the base-tip boundary conditions, leads to a system of equations for the filaments dynamics. Specifically, we have
\begin{equation}
\begin{split}
\mathbf{N}_s^{(\i)}-{\mathbf{f}}^{(\i)}=\mathbf{0},\qquad
\mathbf{M}_s^{(\i)}+ {\r_s}^{(\i)}\times\mathbf{N}^{(\i)}=\mathbf{0}.
\label{eq:eom_filament}
\end{split}
\end{equation}
subject to the free-tip and active-base boundary conditions $(\mathrm{j} = 1,2;\, \mathrm{j}\neq \i)$
%-------
\begin{equation}
\begin{split}
&\mathbf{N}^{(\i)}(\ell,t)  = \mathbf{0}, \qquad \mathbf{M}^{(\i)}(\ell,t) = \mathbf{0},  \qquad
\mathbf{M}^{(\i)}(0,t)  = M^{(\i)} \mathbf{e}_z,  \\ 
&\mathbf{N}^{(\i)}(0,t)\cdot \mathbf{e}_x = K_\c \, \xbi  +K\left(\xbi - \xbj\right), 
\quad \mathbf{r}^{(\i)}(0,t)\cdot\mathbf{e}_z = 0.
\label{eq:bc_filament}
\end{split}
\end{equation}
%----------
The fourth boundary conditions in~\eqref{eq:bc_filament} is derived from a total balance of forces on each filament and basal spring system in the $x$-direction, and the last condition imposes the constraint of no basal motion orthogonal to the $x$-direction.

To fully determine the centerline deformation $\mathbf{r}^{(\i)}(s,t)$ of each filament given the driving moment $M^{(\i)}$ at the filament base, we need to solve the filament system of equations~(\ref{eq:eom_filament},\ref{eq:bc_filament}) coupled to the incompressible Stokes equation
%--------
\begin{equation}\label{eq:stokes}
-\nabla p + \mu\nabla^2\mathbf{u} + \sum_{\i = 1,2} \mathbf{F}^{(\i)} = 0, \qquad \nabla\cdot \mathbf{u}=0.
\end{equation}
%--------
Here, $\mathbf{u}(\mathbf{x},t)$ is the fluid velocity field expressed as a function of the three-dimensional position $\mathbf{x}$ and time $t$, $p(\mathbf{x},t)$ the pressure field, and $\mathbf{F}^{(\i)}(\mathbf{x},t)$ the force per unit volume exerted by filament ${(\i)}$ on the fluid.
We take advantage of the filament slenderness (small aspect ratio $a/\ell\ll 1$)  to write the fluid boundary conditions as $\left. \mathbf{u}^{(\i)}(\mathbf{x},t)\right|_{\rm centerline} = \mathbf{r}_t^{(\i)}(s,t)$ and the hydrodynamic force $\mathbf{F}^{(\i)}(\mathbf{x},t) = \int_{0}^{\ell} \mathbf{f}({s},t) \delta(\mathbf{x} - \mathbf{r}^{(\i)}(s,t)) d{s}$, where $\delta$ is the three-dimensional Dirac delta function.  
We also invoke that, for slender filaments and small inter-filamentous distance $h(s)$ compared to $\ell$, the force $\mathbf{F}^{(\i)}(\mathbf{x},t)$ can be represented by a line of Stokeslets distributed along the centerline of each filament~\cite{Man2016}. Here, we compute the Stokeslet strength and induced velocity field $\mathbf{u}$ numerically using the regularized Stokeslet method~\cite{Cortez2005} as detailed in the electronic supplementary material.

To obtain non-dimensional counterparts to the equations of motion, we consider the dimensional scales associated with the fluid viscosity $\mu$, flagellum length $\ell$, and time scale $\ell^4 \mu /B$ arising from balancing the filament's elasticity and fluid viscosity. The bending rigidity is of the order $B = 800~\text{pN}\cdot\mu\text{m}^2$, as reported in~\cite{Xu2016} for wild type \textit{Chlamydomonas} flagella. A list of the dimensional parameters used to scale the equations of the motion are summarised in Table~\ref{tab:parameter1}. Hereafter, all quantities are considered dimensionless unless otherwise stated.

%------------------------
\begin{table*}[!t]
\begin{center}
\begin{tabular}{lll}
\hline
\textbf{Parameter} & \textbf{Symbol} & \textbf{Value} \\
\hline
Fluid viscosity   &$\mu\approx \mu_{water,20^\circ C}$  & $10^{-3}~\mathrm{Pa}\cdot\mathrm{s}$  \\
Filament length & $\ell$ & $\textcolor{black}{10}~\mu\mathrm{m}$~\cite{Brennen1977, Brokaw1983}\\
Filament bending rigidity &$B$ & $800~\mathrm{pN}\cdot\mu\mathrm{m}^2$~\cite{Xu2016}\\
Time scale & $T \sim {\ell^4 \mu}/{B}$ & \textcolor{black}{$0.0125~\mathrm{s}$} \\
Basal driving moment & ${M}\sim B/\ell$  &$80$--$240~\mathrm{pN}\cdot\mu\mathrm{m}$ \\
Basal inter-filamentous spring stiffness& $K\sim B/\ell^3$  & $8$--$80$ $\mathrm{pN}\cdot{\mu \mathrm{m}}^{-1}$ \\
Basal filament-cell spring stiffness  & $K_\c \sim B/\ell^3$  & $8~\mathrm{pN}\cdot{\mu \mathrm{m}}^{-1}$ \\
\hline
\end{tabular}
\caption{\label{tab:parameter1}Dimensional parameters used in the filament model simulations. 
}
\end{center}
\end{table*}
%------------------------

\section*{Comparison to flagellar synchrony}

To reproduce filament deformations that are comparable to the flagella waveforms observed in \textit{Chlamydomonas reinhardtii}, 
we consider the active moment $M$ to vary nonlinearly with base angle $\theta$ and we assume a configuration-dependent reference curvature $\boldsymbol{\kappa}_o$ and bending stiffness $B$. Specifically, we prescribe two forms of filament actuation: one inspired by the flagella waveforms while beating {inphase} and another by the attenuated waveforms in while beating {antiphase}.
{These waveforms are obtained by considering two amplitudes of the basal switch angle $\Theta$: a large amplitude mimicking the amplitude observed during inphase beating, and a small amplitude mimicking that in antiphase. In both cases we shift the average switch angle $\bar{\Theta}$ away from the neighboring filament, and we let the active moment $M$ be larger during the power stroke; see Figure~\ref{fig:Chlamy}A.}
{More} details of these actuation profiles are given in the electronic supplementary material. Note that the beat frequency is not prescribed a priori and it is an emergent property of the model.

We focus on two identical filaments coupled via basal springs and hydrodynamic coupling.
Depending on the filament actuation profile, the elastic coupling at the bases leads the filaments to exhibit either inphase or antiphase synchrony as shown in Figure~\ref{fig:Chlamy}(C). Inphase synchrony reminiscent of {the one} in Figure~\ref{fig:Chlamy}(B) is obtained at relatively larger values of the basal switch angle $\Theta$. At smaller $\Theta$, with all other parameters unchanged, the filaments reach antiphase synchrony reminiscent of the {one} in Figure~\ref{fig:Chlamy}(B). This attenuation in the range of basal angles also induces faster beating frequency in the {antiphase} mode compared to that {inphase}, consistent with {\em in vivo} observations~\cite{Leptos2013}. 
All simulations in Figure~\ref{fig:Chlamy}C have the same basal spring stiffnesses $K = 15$,  $K_\c = 10$; the distance between the bases is set to  $x^{(1)}-x^{(2)} = 0.25\ell$ when unactuated, and the basal switch amplitudes are set to $\Theta=0.2\pi$ and $\Theta = 0.1\pi$ for inphase and antiphase, respectively. 
The beating frequency of the simulated {antiphase} synchronization is roughly two times faster than that of the {inphase} synchronization. This is a much more pronounced increase when compared to the $\sim50\%$ increase reported experimentally \cite{Leptos2013}.
The overshoot of frequency difference may  have resulted from our idealization of the filament actuation mechanism -- more sophisticated models, e.g. distributed curvature control along the flagella, {might} yield more realistic frequencies.

In addition to the basal angles, the spring force between the bases are also shown in Figure~\ref{fig:Chlamy}(C). 
The spring force of the {inphase} mode is about twice as large as that of the {antiphase} mode, as the bases during the {antiphase} mode often move in the same directions, resulting in a small change of the basal distance (Supplemental Figure 2).

We next examine the case when the actuation of one of the filaments suddenly changes for a short period of time before returning to normal, while the actuation of the other filaments remains unchanged. This scenario is reminiscent to a differential response to an environment cue by the \textit{trans} and \textit{cis} flagella.  Specifically, starting from {inphase} synchrony, we reduced 
the basal switch angle $\Theta$ of one filament from $\Theta = 0.2\pi$ to $0.1\pi$ for a time interval equal to 0.2 dimensionless unit, then set it back to $\Theta = 0.2\pi$.
We observe a phase slip similar to the slip reported in~\cite{Wan2014} and reproduced in Figure~\ref{fig:Chlamy}(B). The two filaments lose their {inphase} synchrony instantaneously as soon as one of the filament's basal switch angle is reduced and recovered synchrony gradually once the angle of the perturbed filament is changed back to its original value.
We note that the time required for the two filaments to re-synchronize depends on the actuation and basal coupling strength, as well as the phase differences when the perturbed filament returns to normal activity. In this particular case the perturbed filament beats one more period than the unperturbed one, and returns to synchrony almost immediately after $\Theta$ is set to its initial value.

\section*{Analysis of filament synchrony}

We consider a simpler actuation model in which the filaments have zero reference curvature $\boldsymbol{\kappa}_o = 0$ and uniform bending stiffness $B$, and where the active moment $M$ and  basal switch angle $\Theta$ are symmetric about the vertical. Thus,
the beating waveforms of each filament are also symmetric about the vertical. 
This simplification retains the essence of the geometric-switch model, and provides a tangible platform for us to analyze the effects of self-actuation via the parameters $M$ and $\Theta$ as well as basal coupling via the parameter $K$ on the synchrony of the filaments.
To distinguish the effect of basal coupling from that of hydrodynamic coupling, which we studied in previous work~\cite{Guo2018,Man2020b}, here we consider only local fluid drag on each filament.
In other words, the no-slip boundary condition along the (i)-th filament's centreline is determined from \eqref{eq:stokes} using $\mathbf{F}^{\rm(i)}$ only, rather than the sum of forces from both filaments.

%---------------------
\begin{figure*}[!t]
        \centerline{\includegraphics[scale=1]{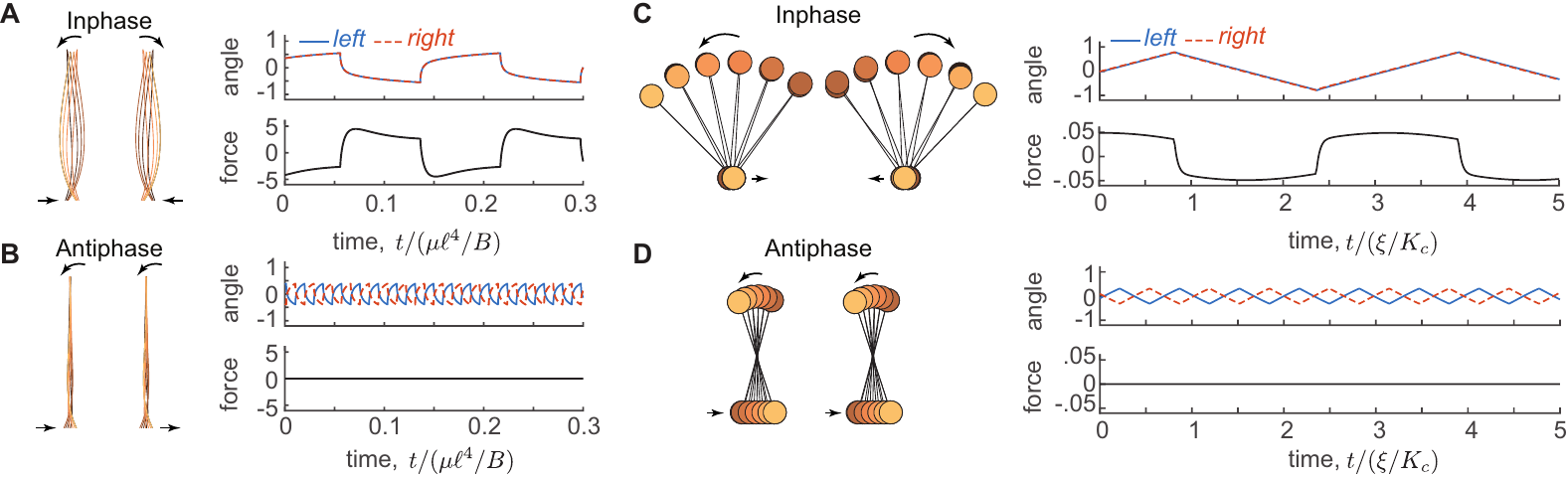}}
	\caption[]{\footnotesize Time evolutions of the filament model and the minimal model. (A, B) Snapshots of filament waveforms and time evolutions of the basal angles $\theta^{(1)}$ and $\theta^{(2)}$ and basal spring force $-K\left(x_\mathrm{b}^{(1)}-x_\mathrm{b}^{(2)}\right)$ showing (A) inphase synchrony for $\Theta= 0.175\pi$, and (B) antiphase synchrony for $\Theta= 0.125\pi$. Other parameter values are set to $K = 50$, $K_\c = 10$, $M = 2$. {The reference curvature of the filament is set to be zero}. 
	(C, D) Snapshots of dumbbell configuration and time evolution of basal angles  and basal spring force  showing (C) inphase synchrony for $\Theta= 0.25\pi$ and (D) antiphase synchrony for $\Theta= 0.1\pi$ synchrony. Other parameter values are set to $K = 20$, $M = 1$.} 
	\label{fig:evolution}
\end{figure*}
%------------------------

The beating waveforms of the two coupled filaments are shown in the left column of Figure~\ref{fig:evolution}(A,B).
The time history of the basal angles $\theta^{(\i)}$ resembles that of the {\em in vivo} results and the flagellum model in Figure~\ref{fig:Chlamy}. When both filaments have the same actuation $M$ and $\Theta$, they synchronize either into {inphase} (Figure~\ref{fig:evolution}A) or {antiphase} (Figure~\ref{fig:evolution}B) depending on the value of $\Theta$, with {inphase} synchronization for larger $\Theta$; perturbing $\Theta$ for one of the filaments for a short time interval leads to a slip (Supplemental Figure~3). 
The basal position $\xbi$ of the filaments also synchronize inphase during {inphase} and antiphase during {antiphase} (Supplemental Figure~3), with the basal distance $x_\mathrm{b}^{(1)}-x_\mathrm{b}^{(2)}$ between the two filaments remaining constant in the {antiphase} mode, leading to zero force in the basal spring $K$ connecting the two filaments, as shown in the bottom row of Figure~\ref{fig:evolution}(B).

%---------------------
\begin{figure*}[!t]
        \centerline{\includegraphics[width=\linewidth]{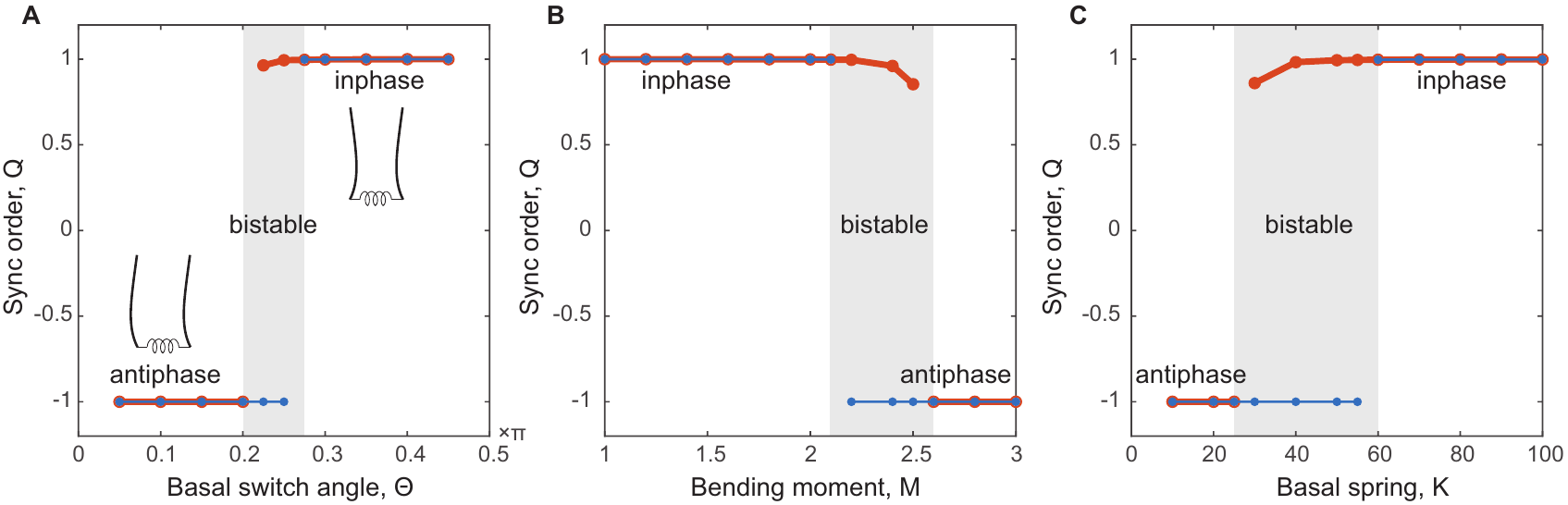}}
	\caption[]{\footnotesize Synchronization order parameter as a function of the filament activity parameters $\Theta, M$ and basal coupling strength $K$, color-coded by the initial condition: simulations with nearly inphase or antiphase initial conditions are depicted in thick red lines or thin blue lines, respectively. 
	The values of the parameters held constant are set to (A)  $M=3, K=50$, (B) $K = 30, \Theta = 0.2\pi$, (C) $M=3, \Theta = 0.25\pi$. Bistable regions are highlighted by shaded boxes.} 
	\label{fig:bistability}
\end{figure*}
%------------------------

We hereafter focus solely on the {inphase} and {antiphase} synchronization modes, and how they are affected by the filament actuation parameters $M$ and $\Theta$ and the basal coupling $K$ between the filaments. 
To quantify the long-term dynamics of the filament pair, we adopt the synchronization order parameter $Q=\int_{T_{n-1}}^{T_n} \text{sgn}(\theta^{(1)}(t)) \text{sgn}(\theta^{(2)}(t))\mathrm{d}t/(T_n - T_{n-1})$~\cite{Kotar2010}, where $T_n$ is the time when the left filament switches from power stroke to recovery stroke for the $n$-th time. 
Specifically, $Q=1$ corresponds to a perfect inphase mode, and $Q=-1$ corresponds to a perfect antiphase mode.
We pick large $n$ (around 100) to ensure that the filaments have settled into their long-term dynamics.

We first examine the effects of the three parameters $M$, $\Theta$,  and $K$ on the long-term dynamics by holding two of the parameters constant while varying the third (Figure~\ref{fig:bistability}).
In all of our simulations, we keep the basal filament-cell spring stiffness $K_c$ a constant that is same as the smallest inter-filamentous spring stiffness $K$ we study.
We use two initial conditions for each parameter combination,
one close to inphase and the other close to antiphase. We find that, in general, filaments with small basal switch angle $\Theta$ or large bending moment $M$ favour antiphase synchrony, \textit{vice versa} for inphase synchrony. On the other hand, stiff basal coupling $K$ favours inphase synchrony. 
Note that there are parameters for which the flagella  synchronize either in {antiphase} or {inphase} mode depending on the initial conditions, meaning that the modes are `bistable'. 
This bistability is similar to what we have observed for hydrodynamic coupling between filaments~\cite{Guo2018,Man2020b}. The results here show that the bistability also exists for two filaments coupled solely via the basal spring.

\begin{figure*}[!t]
        \centerline{\includegraphics[scale=1]{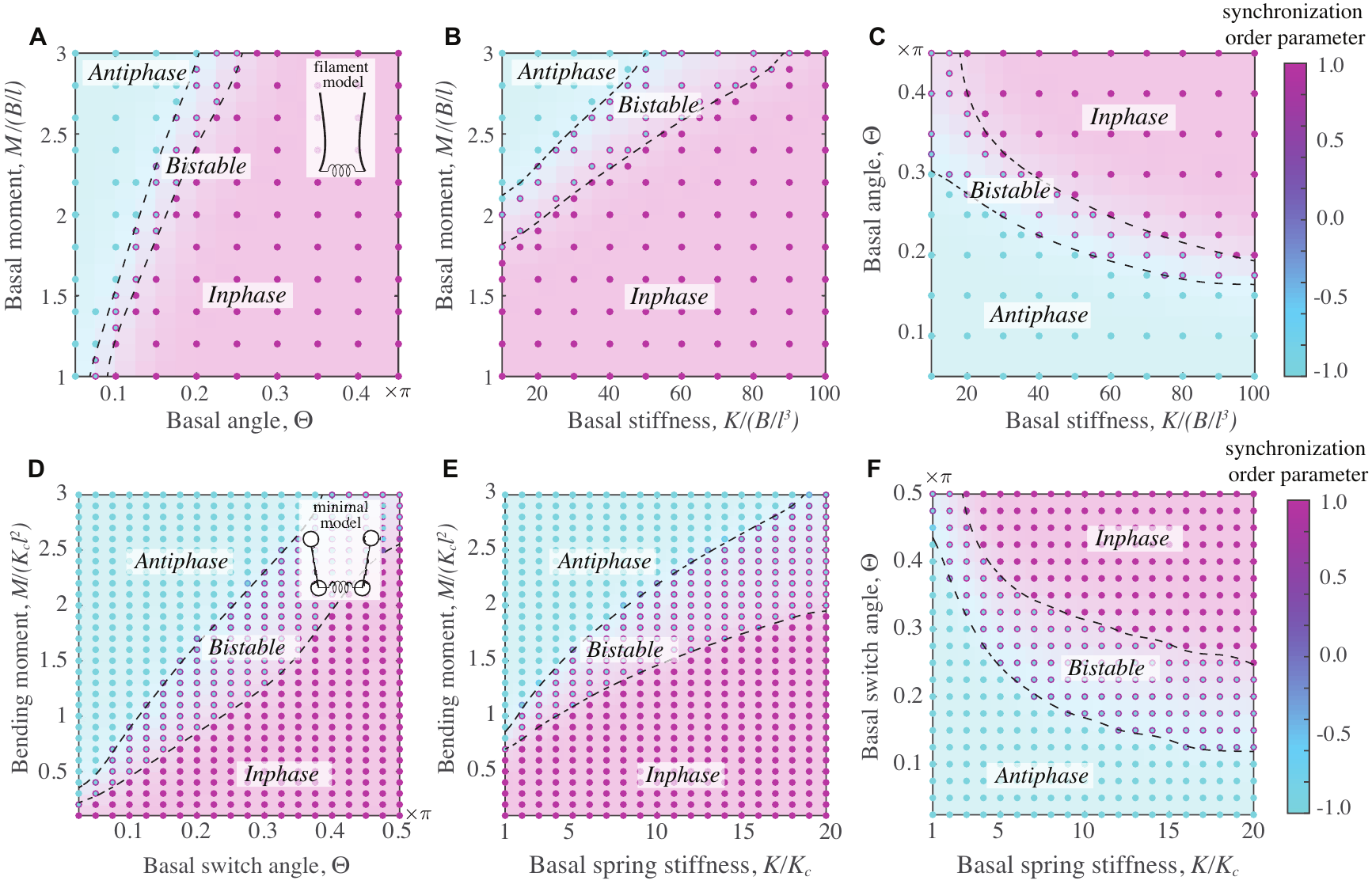}}
	\caption[]{Long-term dynamics of the filament model and the minimal model.
	(A - C): Three slices of the parameter space for the filament model at (A) $K=50$, (B) $\Theta=0.2\pi$, (C) $M=3$ respectively. 
	(D - F): Three slices of the parameter space for the minimal model at (D) $K = 20$, (E) $\Theta = 0.4\pi$, (F) $M = 1$ respectively.
	Each data point comprises the long-term dynamics of two different initial conditions, distinguished by the marker edge and face colors. Flagella synchronized into {inphase} or {antiphase} are in magenta or cyan colors, respectively.
	} 	
	\label{fig:parameterspace}
\end{figure*}

We take the analysis of the long-term dynamics further onto three cross-sections of the three-dimensional parameter space spanned by  $M$,  $\Theta$, and  $K$ (Figure~\ref{fig:parameterspace}A-C).
Similar to Figure~\ref{fig:bistability}, two different initial conditions are used at each point on the cross-sections, one being close to inphase and the other being close to antiphase. 
The results confirm the findings from Figure~\ref{fig:bistability}: while keeping other parameters constant, a larger $M$, a smaller $\Theta$, or a smaller $K$ promotes the synchronization of the two flagella into an {antiphase} mode (cyan color), and the opposite for the {inphase} mode (magenta color).
Physically speaking, the results imply that a softer basal coupling and stronger filament activity lead to {antiphase} synchrony, while a stiffer basal coupling and weaker filament activity lead to {inphase} synchrony.

%---------------------
%------------------------

\section*{Minimal model}

To  examine the fundamental mechanisms underlying synchrony, we consider a minimal dumbbell model consisting of two beads of equal radius $a$ and drag coefficient $\xi = 6 \pi \mu a$, connected via a rigid rod of length $\ell$ and negligible drag.
Similar to the filament model, each dumbbell is actuated at its base by a configuration-dependent moment $M$. A linear elastic spring of stiffness $K_\c$ connects the dumbbell base to a fixed point on the $x$-axis, while two neighboring dumbbells are coupled at their base via an elastic spring of stiffness $K$. 
This simplification maintains two key features of the filament model: the geometric switch actuation and the elastic basal coupling. 

The dynamics of each dumbbell is fully represented by the time evolution of its basal position $x_\b(t)$ and angle $\theta(t)$. In addition to the actuation moment, each dumbbell is subject to spring forces acting at the dumbbell's basal bead, and hydrodynamic drag forces acting on both beads. Hydrodynamic coupling between beads of the same or neighboring dumbbells is neglected. Balance of forces and moments on each dumbbell $(\i=1,2, \ \rm{j} \neq \i)$ leads to the system of equations (see the electronic supplemental material for more details)
%--- 
\begin{equation}
\begin{split}
-K_\c \, \xbi  - K\left(\xbi-\xbj \right)- \xi\left(\dot{x}_\b^{(\i)} + \ell \dot{\theta}^{(\i)}\cos{\theta}^{(\i)}\right) - \xi\dot{x}_\b^{(\i)} & ={0},\\
-\xi\ell\dot{x}_\b^{(\i)} \cos\theta^{(\i)} -\xi\ell^2 \dot{\theta}^{(\i)}  & = M^{(\i)}.
\label{eq:eom_dumbbell}
\end{split}
\end{equation}
%---
We re-write the governing equations in non-dimensional form using  the characteristic length scale $\ell$,  the force scale $\ell K_{\c}$, and  time scale $\xi/K_{\c}$. Further details about the dumbbell model and its non-dimensional form can be found in the electronic supplementary material. Hereafter, all quantities are considered dimensionless unless otherwise stated.

The dynamics of the coupled dumbbell pair is shown in Figure~\ref{fig:evolution}(C,D).
The time history of the  angles $\theta^{(\i)}$ resembles that of the {\em in vivo}  and filament models in Figure~\ref{fig:Chlamy}(B,C) and Figure~\ref{fig:evolution}(A,B). When both dumbbells have the same activity $M$ and $\Theta$, they synchronize either into {inphase} (Figure~\ref{fig:evolution}C) or {antiphase} (Figure~\ref{fig:evolution}D) depending on the value of $\Theta$, with {inphase} synchronization for larger $\Theta$. The basal positions $\xbi$ also synchronize {inphase and antiphase respectively} as noted in the filament model (Supplemental Figure~5), with zero force in the coupling basal spring during {antiphase} synchrony (Figure~\ref{fig:evolution}D). These results imply that, unlike when the synchrony is mediated by hydrodynamic coupling~\cite{Guo2018}, flagellar elasticity is not essential for observing both {antiphase} and {inphase} synchrony when mediated by elastic basal coupling. We explore this further by examining the long-term dynamics of the minimal model over three cross-sections of the parameter space $(M,\Theta,K)$ (Figure \ref{fig:parameterspace}D-F), then over the entire three-dimensional space (Supplemental Figure 6).
The results are qualitatively similar to the filament model: larger $M$, smaller $\Theta$, or smaller $K$ lead to {antiphase} synchrony. This behaviour is not affected by the lack of flagellar compliance in the dumbbell model, emphasizing that the filament elasticity not necessary for obtaining multiple synchronization modes. 
We note that lacking flagellar compliance does not mean that the model has no compliance. In fact, the basal sliding in our model introduces another means of compliance in the system, which has not been studied before.

%%%%%%%%%%%%%%
\subsection*{Mechanism driving synchrony}

There are two time scales of interest in the dimensionless system of the dumbbell model: the relaxation time $T_\b=\xi/K$ dictated by the strength of the basal spring coupling and  the intrinsic oscillation time $T_\a$ dictated by the actuation strength.  
The latter is obtained by balancing the active force $M/\ell$ with the fluid drag $\xi \ell \dot{\theta}$. Setting $\dot{\theta} = \Theta/T_\a$, we arrive at $T_\a = \xi \ell^2\Theta/M$. 
The ratio between $T_\b$ and $T_\a$ is a dimensionless number {\em that measures the relative strength between the flagellar activity and basal coupling}
%---
\begin{equation}
\beta =  \frac{\text{Basal coupling relaxation time}}{\text{Actuation-driven oscillation time}}=\dfrac{T_\b}{T_\a} =  \dfrac{M}{K\ell^2\Theta}.
\label{eq:ratio}
\end{equation}
%---
We propose that this ratio of time scales is predictive of whether the system synchronizes into {inphase} or {antiphase}.
Physically speaking, when $T_\b\gg T_\a$ ($\beta\gg1$), the actuation changes direction much faster than the basal spring can change length, resulting in an almost constant basal spring length, which in turn leads to a {antiphase} synchronization mode.
On the other hand, when $T_b\ll T_a$ ($\beta\ll 1$), the basal spring has enough time to respond to the actuation forces, the basal spring $K$ is thus active and exerts equal and opposite forces at the base. This symmetric basal spring force leads to the (mirror-symmetric) {inphase} synchrony.

\begin{figure*}[!t]
        \centerline{\includegraphics[scale=1]{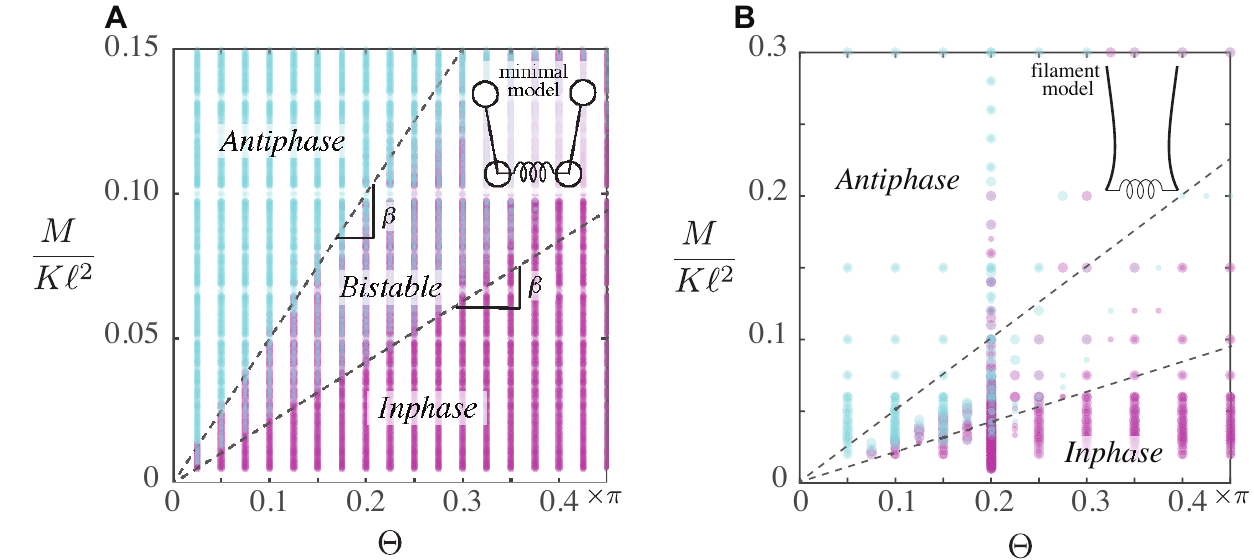}}
	\caption[]{\footnotesize (A) Synchronization modes of dumbbell model plotted in the condensed 2D parameter space $(\Theta, M/K\ell^2)$. Dash lines delineate boundaries between different regions by their stable synchronization modes. (B) Synchronization modes of filament model plotted in the same parameter space. The dash lines in B have the same slopes as those in A.} 
	\label{fig:scaling}
\end{figure*}

To test our proposition, we examine the synchronization modes over the entire parameter space $(M,\Theta, K)$ of the dumbbell model (Supplemental Figure 6), and we project these results onto a two-dimensional parameter space, spanned by the two dimensionless parameters $\Theta$ and $M/K\ell^2$ (Figure~\ref{fig:scaling}A). 
The slope of the line connecting the origin and any point on the 2D parameter space is thus the corresponding $\beta$ of that point, measuring the actuation strength relative to the strength of the basal coupling.
These results demonstrate that the parameter space can be {loosely} divided into three regions with distinct synchronization modes. {Dashed lines are sketched on top of the data as visual guidance.} Specifically, a stable {antiphase} region ($\beta>0.16$) and stable {inphase} region ($\beta<0.067$) are identified, where almost all cases synchronize into {antiphase} or {inphase} mode respectively, regardless of the initial condition. Additionally, there is a bistable region for intermediate $\beta$, where the synchronization modes depend on the initial conditions. 

Finally, to check this scaling law for the filament model, we project the results of the filament synchronization from the three cross-sections of the 3D parameter space (Figure~\ref{fig:parameterspace}A-C) onto the same 2D parameter space (Figure~\ref{fig:scaling}B). Again, we identify regions with distinct {inphase} and {antiphase} synchronizations. A `mixed' region is also observed where either {inphase} or {antiphase} could emerge depending on the model parameters and initial conditions. The delineation between these regions is nonlinear at large values of $M/K\ell^2$. 
We speculate this nonlinearity is due to the compliance of the filament (finite bending rigidity $B$), which introduces another time scale that is not well-captured by the scaling law in~\eqref{eq:ratio}.
In fact,  we show that the scaling law in~\eqref{eq:ratio} holds, modulo a constant factor, in the limit of stiff filament ($B\rightarrow\infty$) as detailed in the electronic supplemental material. 
Here, we superimpose two dashed lines of the same slope as those in the minimal model onto the filament model in Figure~\ref{fig:scaling}(B); although the lines do not form definite boundaries between the three domains at large $M/K\ell^2$, they are indicative of the synchronization state in a rough sense: $\beta<0.067$ indicates {inphase} and $\beta>0.16$ is mostly {antiphase}.

%---------------------
%------------------------

\section*{{Discussion}}

We showed in the context of an \textit{in-silico} filament model, that two flagella coupled at their base via elastic basal springs, with no hydrodynamic interactions, can reach multiple synchronization modes. Both {inphase} and {antiphase} synchrony can be achieved by modulating either the filament activity level or the elastic stiffness of the basal coupling or both. Neither hydrodynamic coupling nor flagellar compliance are necessary to reach these synchronization modes. {Antiphase} synchrony is characterized by higher beating frequencies, as noted experimentally. 
Perturbing the basal switch angle of one flagellum for a short period of time reproduces the phase slip mode observed in wild type \textit{Chlamydomonas}. The time duration of a slip could well be environment-dependent~\cite{Wan2014}, and this model allows us to alter the transition dynamics by changing the perturbing time interval.
The time required for synchrony to re-establish following a perturbation (after the basal switch angle is changed back to its original value) is determined by the filament actuation and basal coupling parameters, together with the instantaneous phase difference. Specifically, a stiffer basal coupling spring would re-synchronize the two filaments faster than a softer one.

To further analyze transitions between {inphase} and {antiphase} synchrony, we introduced a dimensionless parameter $\beta= T_\b/T_\a$, which is the ratio of two time scales: a time scale $T_\b$ that arises from balancing the hydrodynamic drag forces with basal spring forces, and a time scale $T_\a$ from balancing drag with the actuation forces (flagellar activity). 
Loosely speaking, $\beta$ measures the relative strengths between flagellar activity and stiffness of basal coupling. When flagellar activity is dominant, the basal connection is overwhelmed and the filaments synchronize into {antiphase}. When flagellar activity is moderate and the basal spring has sufficient time to respond via elastic forces to the filament motion, it exerts equal and opposite forces at the filaments' base, thus driving them into {inphase} synchrony.

We verified that this scaling analysis is predictive of flagellar synchrony in the context of our minimal dumbbell model, and it maps reasonably well to the filament model. Based on our results (Figure~\ref{fig:scaling}), $\beta < 0.067$ is indicative of {inphase} synchrony and $\beta > 0.16$ of {antiphase}.
In addition to its utility in predicting the flagella synchronization modes, this analysis could be used to provide estimates on currently hard-to-measure quantities, such as the flagellar activity level.
For example, knowing the length $\ell =10\mu\mathrm{m}$ of the \textit{Chlamydomonas} flagella and that the amplitude $\Theta$ of the basal switch angles   is roughly equal to $0.6$ rad and $0.5$ rad during {inphase} and {antiphase} respectively~\cite{Leptos2013}, and assuming the bending moment is constant for both {inphase} and {antiphase}, the threshold for $\beta$ implies that the basal spring stiffness during the {inphase} mode would be one and half times as stiff as that during the {antiphase} mode. 
Conversely, if we assume the basal spring stiffness to be $K=100$ pN/$\mu$m~(Table~\ref{tab:parameter1}), we can estimate the active moment at the base to be $M\approx 0.07K\ell^2\Theta=420\mathrm{pN}\cdot\mu\mathrm{m}$ for {inphase}, and $M\approx 0.16K\ell^2\Theta=800\mathrm{pN}\cdot\mu\mathrm{m}$ for {antiphase}. 
{Note that neither $M$ nor $K$ is directly accessible via state-of-the-art imaging and image-processing methods. Our analysis provides a simple connection between these two parameters: provided that we know the mode of synchrony, if we can measure one parameter, we can estimate the other.}

Our findings have important biological implications for the active control of flagellar coordination, and demonstrate a concrete mechanism by which cells can directly control and manipulate the synchronization state of their flagella in real-time. Flagellated cells have a great incentive for efficient swimming that is robust to noise and hydrodynamic perturbations, and also for reliable transitions between swimming gaits, such as between forward swimming and turning. These gaits and transitions are mediated by flagellar synchrony. In the \textit{Chlamydomonas} example, robust forward swimming is associated with {inphase} synchrony, and turning occurs during asynchronous beating or transition to {antiphase}. 
In different species of quadriflagellates, changes in the activity level of individual flagella can influence the global synchronization state of the coupled flagella network \cite{Wan2020}. 
These gait transitions occur stochastically, and at rates sensitive to environmental factors. Ceding the control of flagellar synchrony to hydrodynamic coupling seems to be too costly for an organism with relatively few flagella. It would imply little or no control by the cell over its swimming gait or its switching rate.

{Our model explicitly accounts for the marked change in the beating patterns reported \textit{in vivo}, which provides a way to assess the effect of beat form on synchrony.}
Given the similarities between the {antiphase} flagellar beat pattern in \textit{ptx1} and that of the faster flagellum during wildtype phase slips (Figure 1B,C), our results support the notion that change in filament activity could \textit{initiate} transitions in synchronization state, rather than the reverse. 
Thus, biochemical fluctuations of a similar nature could induce phase-slips in the wildtype yet antiphase episodes in \textit{ptx1} - which nominally have two \textit{trans}-flagella \cite{Leptos2013}. 
Such distinctions highlight the subtle functional differences existing between the two outwardly-similar \textit{Chlamydomonas flagella}, which differ primarily in the generational age of their basal bodies.

{Our results support the hypothesis that a flagellated cell, with intact basal connections, could control flagellar synchrony in two ways: (i) by modulating the level of flagellar activity, with little or no change to the contractile properties of the basal connections, or (ii) by modulating the contractility of these basal connections, while sustaining the same level of flagellar activity. In reality, it is likely that both filament activity level and basal contractility depend on common signalling pathways or abundance of the same molecule (e.g. ATP). 
For example, Ca$^{2+}$ is a ubiquitous second messenger of cilia in many organisms from protists to metazoa \cite{Silflow2001,Satir2007}. 
Calcium ions not only control the beating waveform and frequency of isolated flagella axonemes \cite{Bessen1980}, 
but also the contractile state of centrin - a key cytoskeletal protein and constitutent of algal basal apparatuses \cite{Salisbury1989,Wan2016}.
At the biomolecular level, calcium binding and detachment likely alters the entropic state of centrin-type biopolymers \cite{Moriyama1999} in the basal apparatus. 
Real-time reorientation of the V-shaped basal apparatuses of \textit{Chlamydomonas} has also been observed in response to changing Ca$^{2+}$ \cite{Hayashi1998}. 
Further insights into these processes in live cells will certainly require additional modelling in parallel with experimentation.}

\section*{Acknowledgements}
The authors would like to acknowledge useful conversations with Janna Nawroth and Michael J. Shelley. The work of HG, YM and EK is partially supported by the National Science Foundation (NSF) INSPIRE grant 1608744 and the University of Southern California (USC) Bridge fund. KYW acknowledges funding from the European Research Council under the European Union's Horizon 2020 research and innovation programme (grant agreement no. 853560).

\bibliographystyle{unsrt}
\bibliography{reference}

\end{document}